\documentclass[5p,twocolumn ]{elsarticle}   
\usepackage{bbm}
\usepackage{graphicx}
\usepackage{bm}
\usepackage{amsmath}
\newcommand{\cN}{{\cal N}}
\newcommand{\cD}{{\cal D}}
\newcommand{\cP}{{\cal P}}

\DeclareMathOperator{\Det}{Det}
\DeclareMathOperator{\Tr}{Tr}
\DeclareMathOperator{\tr}{tr}

\newcommand{\be}{\begin{equation}}
\newcommand{\ee}{\end{equation}}

\newcommand{\bal}{\begin{align}}
\newcommand{\eal}{\end{align}}

\newcommand{\rk}{\right)}
\newcommand{\lk}{\left(}

\newcommand{\ii}{\mathrm{i}}
\newcommand{\e}{\text{e}}

\renewcommand*{\vec}[1]{\bm{\mathrm{#1}}}

\newcommand{\vB}{\vec{B}}
\newcommand{\ve}{\vec{e}}
\newcommand{\vA}{\vec{A}}
\newcommand{\vx}{\vec{x}}
\newcommand{\vy}{\vec{y}}
\newcommand{\vD}{\vec{D}}
\newcommand{\vp}{\vec{p}}
\newcommand{\va}{\vec{a}}
\newcommand{\vd}{\vec{d}}
\newcommand{\vz}{\vec{z}}
\newcommand{\vC}{\vec{C}}
\newcommand{\vq}{\vec{q}}
\newcommand{\mR}{\mathbbm{R}}

\newcommand{\bra}[1]{\langle #1\rvert}
\newcommand{\ket}[1]{\lvert#1\rangle}

\newcommand*{\vev}[1]{\left< #1 \right>}

\newcommand{\kvec}[1]{\mbox{\boldmath$\scriptstyle{#1}$\unboldmath}}

\newcommand{\kvA}{{\kvec{A}}}

\newcommand{\il}{\int\limits}
\newcommand{\sli}{\sum\limits}
\newcommand{\cL}{{\cal L}}

\newcommand*{\dd}[1][]{\mathop{\mathrm{d}^{#1}}\mkern-4mu}

\begin{document}

\title{The effective potential of the confinement order parameter in the Hamilton approach}

\author{Hugo~Reinhardt}
\author{Jan~Heffner}

\address{Institut f\"ur Theoretische Physik,\,Auf der Morgenstelle 14,\,D-72076 T\"ubingen,\,Germany}

\begin{abstract}
The effective potential of the order parameter for confinement is calculated within the Hamiltonian approach
by compactifying one spatial dimension and using a background gauge fixing. Neglecting the ghost and using the 
perturbative gluon energy one recovers the Weiss potential. From the full non-perturbative potential calculated 
within a variational approach a critical temperature of the deconfinement phase transition of $269$ MeV is found for 
the gauge group SU$(2)$. 
\end{abstract}


\maketitle

\section{Introduction}

Understanding the phase structure of QCD is one of the major challenges of particle physics \cite{BraunMunzinger:2009zz}. Running and upcoming
high-energy heavy-ion experiments call for a deeper understanding of hadronic matter under extreme conditions. The
central issue is the understanding of the deconfinement phase transition from the confining hadronic phase with
chiral symmetry spontaneously broken to the deconfined quark-gluon plasma with chiral symmetry restored. This transition
is expected to be driven by the gluon dynamics and lattice calculations show that confinement is exclusively determined
by the strongly interacting low-energy gluonic modes \cite{Yamamoto:2009de}. Therefore, understanding the deconfinement phase transition
requires non-perturbative methods. In quenched QCD reliable results on the deconfinement phase transition have been 
obtained by means of the lattice approach \cite{Karsch:2001cy}, which, however, fails at large baryon densities due to the notorious 
fermion sign problem. Therefore, alternative non-perturbative methods based on the continuum formulation of QCD are
required. In recent years substantial progress has been achieved within continuum approaches to QCD
 \cite{Fischer:2008uz,Pawlowski:2010ht,Binosi:2009qm,Feuchter:2004mk}. Among these
is the variational approach to the Hamiltonian formulation of QCD in Coulomb gauge \cite{Feuchter:2004mk,Epple:2006hv} 
(see also Refs.~\cite{Schutte:1985sd,Szczepaniak:2001rg} for related
earlier work). In this approach the energy density is minimized using Gaussian type ans\"atze for the Yang--Mills vacuum
wave functional. Within this approach a decent description of the infrared sector of Yang--Mills theory was obtained 
\cite{Schleifenbaum:2006bq,CamRei08,Reinhardt:2007wh,Rei08}.
Recently, this approach was extended to finite temperatures by considering the grand canonical ensemble making a 
suitable quasi-particle ansatz for the density operator and minimizing the free energy \cite{Reinhardt:2011hq,HefReiCam12}. In this letter we present
an alternative Hamiltonian approach to finite temperature Yang--Mills theory, which does not require an ansatz for 
the density operator. The finite-temperature is introduced here by compactifying one spatial dimension.

\section{Order parameter for confinement}
\label{sectionII}

As is well known, Euclidean quantum field theory can be extended to finite temperature $L^{- 1}$ by compactifying the Euclidean 
time dimension to an effective length $L$. At temperature $L^{- 1}$ the order parameter for confinement is the expectation value $\langle P [A_0] \rangle$
of the Polyakov loop \cite{Svetitsky:1982gs} ($\cP$ path-odering)
\be
\label{1}
P [A_0] = \frac{1}{N} \tr \cP \e^ { - \int^L_0 \dd x^0 A_0 \lk x^0, \vx \rk} \, .
\ee
This quantity is related to the free energy of a static (infinitely heavy) quark at spatial position $\vx$. In the absence
of fermions Yang--Mills theory is invariant under gauge transformations $U (x^0, \vx) \in \text{SU} (N)$ being periodic up to
a center element $z_k \in \text{Z} (N)$
\be
\label{2}
U (L, \vx) = z_k U (0, \vx) \, .
\ee
Under such a gauge transformation the Polyakov loop transforms as $P [A^U_0] = z_k P [A_0]$ and as a consequence 
$\langle P [A_0] \rangle = 0$ in the center symmetric confining phase while $\langle P [A_0] \rangle \neq 0$
in the deconfining phase with center symmetry spontaneously broken. In Polyakov gauge, $\partial_0 A_0 = 0$,
and with $A_0$ residing in the Cartan algebra, in the fundamental modular region $P[A_0]$ is a convex function of $A_0$ and by Jensen's inequality $\langle P [A_0] \rangle \leq P [\langle A_0 \rangle ]$,
instead of $\langle P [A_0] \rangle$, one can alternatively use $P [\langle A_0 \rangle ]$  or $\langle A_0 \rangle$
as order parameter for confinement \cite{Marhauser:2008fz,Braun:2007bx}. 
Note however, that by gauge invariance a non-vanishing $\langle A_0 \rangle$ requires the presence of an external background
field $a_0$. Choosing $a_0$ in Polyakov gauge and to satisfy $\langle A_0 \rangle = a_0$ the background field becomes 
an order parameter for confinement whose value is determined by the minimum of the effective potential $V \left[
\langle A_0 \rangle = a_0 \right]$.
In Polyakov gauge there are still residual gauge transformations satisfying Eq.~(\ref{2}), which transform $A_0$ to $A^U_0 = A_0 + \mu_k / L$, where $\mu_k$ is  a coweight satisfying $\exp (- \mu_k) = z_k$, 
and as a consequence of gauge invariance the effective potential of $\langle A_0 \rangle = a_0$ must obey the periodicity 
condition 
\be
\label{3}
V \left[ a_0 + \mu_k / L \right] = V \left[ a_0 \right] \, .
\ee
This potential was first calculated in Ref.~\cite{Weiss:1980rj} in one-loop perturbation
theory. It was found that $V [a_0]$ is minimal at $a_0 = 0$, so that $\langle P \rangle \simeq P\left[ \langle A_0 
\rangle = 0 \right] = 1$ implying that the perturbative theory is in the non-confining phase with center symmetry 
broken. This, of course, is the expected behavior at high temperatures, where perturbation theory is reliable. In this letter
we calculate non-perturbativley the effective potential $V [\langle A_0 \rangle]$ in the Hamiltonian approach and
determine from this potential the critical temperature of the deconfinement phase transition.

\section{Finite temperature from compactification of a spatial dimension}
\label{sec3}
Clearly the order parameter $\langle P \rangle \approx P [ \langle A_0 \rangle ]$ or $\langle A_0 \rangle$ is not
directly accessible in Weyl gauge $A_0 = 0$, which is assumed in the canonical quantization. However, by O$(4)$
symmetry, all four Euclidean dimensions are equivalent and instead of compactifying the time, one can equally well 
introduce the temperature by compactifying one of the spatial dimension, say the $x_3$-axis, and consider $\langle A_3 \rangle$
as order parameter for confinement. 
This can be seen as follows:

Consider Yang-Mills theory at finite temperature $L^{- 1}$, which is defined by the partition function
\be
\label{en-1}
Z (L) = \Tr \e^{- LH (\kvA)} \, .
\ee
Here $H (\vA)$ is the usual Yang-Mills Hamiltonian defined by canonical quantization in Weyl gauge $A^0 = 0$.
The partition function (\ref{en-1}) can be equivalently represented by the Euclidean functional integral, see 
for example Ref.~\cite{Reinhardt:1997rm}
\be
\label{en2}
Z (L) = \il_{x^0 - pbc} \prod\limits_\mu \cD A_\mu (x) \, \e^{- S [A]} \, ,
\ee
where
\be
\label{en-3}
S [A] = \il^{L/2}_{- L/2} \dd x^0 \int \dd[3] x \, \cL \lk A^\mu ; x^\mu \rk
\ee
is the Euclidean action and the functional integration is performed over temporally periodic fields
\be
\label{en-190}
A^\mu \lk \frac{L}{2}, \vx \rk = A^\mu \lk - \frac{L}{2}, \vx \rk \, ,
\ee
 which is
indicated in Eq.~(\ref{en2}) by the subscript $x^0 - pbc$. This boundary condition is absolutely necessary at finite
$L$ but becomes irrelevant in the zero temperature $(L \to \infty)$ limit. 

We perform now the following change of variables 
\begin{align}
\label{en-5}
x^0 \to z^3 & \quad \quad A^0 \to C^3 \nonumber\\
x^1 \to z^0 & \quad \quad A^1 \to C^0 \nonumber\\
x^2 \to z^1 & \quad \quad A^2 \to C^1 \nonumber\\
x^3 \to z^2 & \quad \quad A^3 \to C^2 \, .
\end{align}
Due to the $O (4)$ invariance of the Euclidean Lagrangian we have 
\be
\label{en-6}
\cL \lk A^\mu, x^\mu \rk = \cL \lk C^\mu, z^\mu \rk
\ee
and the partition function (\ref{en2}) can be rewritten as 
\be
\label{en-7}
Z (L) = \il_{z^3 - pbc} \prod_\mu \cD C^\mu (z) \, \e^{- \tilde{S} [C^\mu]} \, ,
\ee
where the action is now given by
\be
\label{en-8}
\tilde{S} [C^\mu] = \int \dd z^0 \dd z^1 \dd z^2 \il^{L/2}_{- L/2} \dd z^3 \cL \lk C^\mu, z^\mu \rk 
\ee
and the functional integration runs over fields satisfying periodic boundary condition in the $z^3-$direction
\be
\label{en-9}
C^\mu \lk z^0, z^1, z^2, L/2 \rk = C^\mu \lk z^0, z^1, z^2, - L/2 \rk \, .
\ee
We can now interprete $z^0$ as time and $\vz = \lk z^1, z^2, z^3 \rk$ as space coordinates and perform a usual canonical 
quantization in ``Weyl gauge'' $C^0 = 0$, interpreting $\vC = \lk C^1, C^2, C^3 \rk$ as spatial coordinates of the gauge field, which
are defined, however, not on $\mR^3$ but instead on $\mR^2 \times S^1$. We obtain then the usual Yang-Mills Hamiltonian in which, however, the integration over $z^3$ is restricted to the intervall $\left[- \frac{L}{2}, \frac{L}{2} \right]$. Let us denote this Hamiltonian by $\tilde{H}(\vC,L)$. Obviously $\tilde{H}(\vC ,L \to \infty ) =H(\vC)$.
Reversing the steps which lead from (\ref{en-1}) to (\ref{en2}) and taking into account the irrelevance of the temporal 
boundary conditions in the functional integral
for an infinite time-interval we obtain from Eq.~(\ref{en-7}) the alternative representation of the partition function
\be
\label{en-232}
Z (L) = \Tr \e^{- \int \dd z^0 \tilde{H}(\vC,L)} = \lim\limits_{T \to \infty} \Tr \e^{- T \tilde{H}(\vC,L)} \, .
\ee
Due to the infinite $z^0$-(time-)interval $T \to \infty$ only the lowest eigenvalue of $\tilde{H}(\vC,L)$ contributes to the partition 
function $Z (L)$. The calculation of $Z (L)$ is thus reduced to solving the Schr\"odinger equation $\tilde{H}(\vC,L)\psi (\vC)= E \psi (\vC)$ for the vacuum state on the space manifold $\mR^2 \times S^1 (L)$, where $S^1 (L)$ is a circle with circumference $L$. 

Let us illustrate the equivalence between Eqs.~(\ref{en2}) and (\ref{en-232}) by means of the free scalar field theory in $1 + 1$ dimension defined by the 
(Euclidean) Lagrangian
\be
\label{242-x1}
\cL  = \frac{1}{2} \lk \partial_\mu \phi \rk^2 + \frac{m^2}{2} \phi^2 \, .
\ee
Calculating the partition function for this model from the functional integral (\ref{en2}) with the temporally periodic boundary condition
$\phi (L/2) = \phi (- L/2)$ one finds
\begin{align}
\ln Z (L) &= - \frac{1}{2} \Tr \ln \lk - \partial^2 + m^2 \rk \nonumber\\
&= - \frac{1}{2} \il^L_0 \dd x^0 \int \dd x^1 \frac{1}{L} \sli_n \int \frac{\dd p}{2 \pi} \ln \lk p^2_n + p^2 + m^2 \rk \, ,
\end{align}
where the $p_n = 2 \pi n / L$ are the usual Matsubara frequencies. Representing the logarithm by a proper-time integral, carrying out the 
integral over the spatial momentum $p$ and using the proper-time representation of the square root one obtains 
\be
\label{254-x3}
\ln Z (L) = - \int \dd x^1 E_0 (L) \, ,
\ee
where
\be
\label{259-x4}
E_0 (L) = \frac{1}{2} \sum_n \sqrt{p^2_n + m^2}
\ee
is identified as the ground state energy (the lowest eigenvalue of the corresponding Hamiltonian) of the 
scalar field theory (\ref{242-x1}) defined, however, on a compact spatial manifold $S^1 (L)$. With the
substitution (\ref{en-5}) $x^1 \to z^0$ Eq.~(\ref{254-x3}) is precisely the representation (\ref{en-232}).

The upshot of the above consideration is that finite temperature gauge theory can be described in the Hamiltonian approach
by compactifying a spatial dimension and solving the corresponding Schr\"odinger equation for the vacuum sector. 
This equivalence holds in fact for any $O (4)$ invariant quantum field theory.

The above consideration for the partition function can be extended to the finite temperature effective potential $V [\langle A_0 \rangle]$. One finds
that $V [\langle A_0 \rangle = a]$ can be calculated in the Hamiltonian approach from $V [\langle A_3 \rangle = a]$ with the $z^3$-axis
compactified. Furthermore, as shown in Ref.~\cite{WeinbV2}, in the Hamilton approach the effective potential $V [\langle A_3 \rangle = a]$ is
given by the energy density in the state minimizing $\langle H \rangle$ for given $\langle A_3 \rangle$. 

Below we calculate the effective potential $V \left[ \langle A_3 \rangle = a_3 
\right]$ in the Hamiltonian approache exploiting the representation (\ref{en-232}) of the partition function.

\section{Hamiltonian approach in background gauge}

In the presence of an external constant background field $a$ the Hamiltonian approach turns out to be most conveniently 
formulated in the background gauge
\be
\label{4}
\hat{\vd} \cdot \vA = 0 \,, \quad \hat{\vd} = \vec{\partial} + \hat{\va} \, , \quad \hat{\va}^{ab} = f^{acb} \va^c \, ,
\ee
where the hat ``\, $\hat{\phantom{}}$\, '' denotes the adjoint representation. This gauge allows for an explicit 
resolution of Gauss' law, which results in the gauge fixed Yang--Mills Hamiltonian
\be
\label{5}
H = \frac{1}{2} \int \dd[3] x \lk J_A^{- 1} \vec{\Pi} (\vx) J_A \cdot \vec{\Pi} (\vx) + \vB^2 (\vx) \rk + H_\text{C} \, ,
\ee
where $\Pi^a_k (\vx) = -  \ii\delta / \delta A^a_k (\vx)$  is the ``transversal''
momentum operator ($\hat{\vd} \cdot \vec{\Pi} = 0$) and
\be
\label{6}
J_A = \Det \lk - \hat{\vD} \cdot \hat{\vd} \rk \,, \quad \hat{\vD} = \vec{\partial} + \hat{\vA}
\ee
is the Faddeev-Popov determinant of the gauge (\ref{4}). Furthermore,
\be
\label{7}
H_\text{C} = \frac{g^2}{2} \int \dd[3] x \dd[3] y\,J_A^{- 1} \, \rho^a (\vx) J_A \, F^{ab} (\vx, \vy) \rho^b (\vy)
\ee
arises from the kinetic energy of the ``longitudinal'' part of the momentum operator. Here
\be
\label{8}
\rho^a = - \hat{\vD} \cdot \vec{\Pi} = - \lk \hat{\vA} - \hat{\va} \rk \cdot \vec{\Pi}
\ee
is the color charge density of the gluons, which interacts through the kernel 
\be
\label{9}
F = \lk - \hat{\vD} \cdot \hat{\vd} \rk^{- 1} \lk - \hat{\vd} \cdot \hat{\vd} \rk \lk - \hat{\vD} \cdot \hat{\vd} \rk^{- 1} 
\, .
\ee
For a vanishing background field $a = 0$ the gauge (\ref{4}) reduces to the ordinary Coulomb gauge and $H$ 
(\ref{5}) becomes the familiar Yang--Mills Hamiltonian in Coulomb gauge \cite{ChrLee}.

We are interested here in the energy density in the state $\psi_a [A]$ minimizing $\vev{ H }_a = \bra{\psi_a } H \ket{\psi_a }$
under the constraint $\vev{A}_a = a$. For this purpose we perform a variational calculation with the trial
wave functional 
\be
\label{10}
\psi_a [A]  =  J_A^{- 1/2} \tilde{\psi} [A - a]\,,\,\,
\tilde{\psi} [A]  = \cN \e^{\,\left[ - \frac{1}{2} \int A \omega 
A \right]} \, ,
\ee
which already fulfills the constraint $\vev{A } = a$. For $a = 0$ this ansatz reduces to the trial wave functional used
in Coulomb gauge \cite{Feuchter:2004mk}. However, due to the presence of the colored background field the variational kernel $\omega (\vp)$
is now a non-trivial color matrix. Proceeding as in the variational approach in Coulomb gauge \cite{Feuchter:2004mk}, from 
$\langle H \rangle_a \to min$ one derives a set of coupled equations for the gluon and ghost propagators
\begin{align}
\label{11}
\cD &= \langle A A  \rangle_0 = \frac{1}{2} \omega ^{-1}\,,\quad G =  
- \vev{ \lk  ( \hat{\vec{D}} + \hat{\va} ) \hat{\vd} \rk^{- 1} }_0 \, .
\end{align}
Using the same approximation as in Ref.~\cite{HefReiCam12} in Coulomb gauge, i.e. restricting to two loops in the energy, while neglecting $H_\text{C}$ (\ref{7}) and also the tadpole arising from the non-Abelian part of the magnetic energy, one finds from the
minimization of $\langle H \rangle_a$ the gap equation 
\be
\label{12}
\omega^2 =   - \hat{\vd} \cdot \hat{\vd}  + \chi^2
\ee
where\footnote{We use here the compact notation $A (1) \equiv A^{a_1}_{i_1} (\vx_1)$. For Lorentz scalars like the ghost, the index ``$1$'' stands for 
the color index $a_1$ and the spatial position $\vx_1$.~Repeated indices are summed/integrated over.}
\be
\label{13}
\chi (1, 2) = - \frac{1}{2} \vev{ \frac{\delta^2 \ln J [A + a]}{\delta A (1) \delta A (2)} }_0 
= \frac{1}{2} \Tr \left[ G \Gamma (1) G \Gamma_0 (2) \right]
\ee
is the ghost loop (referred to as ``curvature'') with $\Gamma_0$ and $\Gamma$ being the bare and full ghost-gluon vertex. 
The gap equation (\ref{12}) has to be solved together with the Dyson-Schwinger equation (DSE) for the ghost propagator
\be
\label{14}
G^{- 1} = - \hat{\vd} \cdot \hat{\vd} - \Gamma_0 (1) G \Gamma (2) \cD (2, 1)\,.
\ee
Due to the presence of the background field these equations have a non-trivial color structure. Fortunately, due to the choice
of the background gauge (\ref{4}), the background field enters these equations only in form of the covariant derivative $\hat{d} = \partial + \hat{a}$.
Choosing the background field in the Cartan algebra the above equations can be diagonalized in color space. For simplicity, let us
consider the gauge group SU$(2)$ so that $\hat{a} = a \hat{T}_3$ (the extension to SU$(N)$ is straightforward). The eigenvectors $\ket{\sigma =0, \pm 1}$
of $\hat{T}_3 ^{ab} = \varepsilon^{a 3 b}$ are the spin-1 eigenstates, see e.g.~\cite{ReiSch09}. 
Since the explicit color dependence is only due to the background field 
$a$ the various propagators have to become diagonal in the basis which diagonalizes $\hat{a}$. 
Indeed, one can show that the above equations (\ref{12}) and (\ref{14}) can be consistently solved for propagators of the form
(in momentum space) 
\be
\label{17}
\cD^{\sigma \tau} (\vp) = \delta^{\sigma \tau} \cD^\sigma (\vp) \,, \quad G^{\sigma \tau} (\vp) = \delta^{\sigma \tau} G^\sigma (\vp) \, .
\ee
In addition, one can show that the propagators $\cD^\sigma (\vp)$, $G^\sigma (\vp)$ are related to the propagators in Coulomb gauge
in the absence of the background field, $\cD (\vp)$, $G (\vp)$, by 
\be
\label{18}
\cD^\sigma (\vp) = \cD (\vp^\sigma) \,, \quad G^\sigma (\vp) = G (\vp^\sigma) \, ,
\ee
where 
\be
\label{19}
\vp^\sigma = \vp - \sigma \va
\ee
is the momentum shifted by the background field. The relation (\ref{18}) implies $\chi^\sigma(\vp)=\chi(\vp^\sigma)$ and applies also to the transversal projector $t_{kl}(\vp)= \delta_{kl} - p_k p_l$ of the
Lorentz tensors in Coulomb gauge
\be
\label{20}
\cD_{kl} (\vp) = t_{kl} (\vp) \frac{1}{2 \omega (\vp)} \, , \quad \chi_{kl} (\vp) = t_{kl} (\vp) \chi (\vp) \,.
\ee
\begin{figure}[t]
 \includegraphics[width=0.8\linewidth]{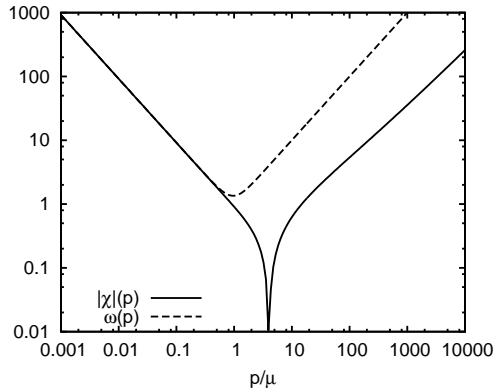}
\caption{The gluon energy $\omega(p)$ and the curvature $\chi(p)$ resulting from the full numerical solution of
the variational approach in Coulomb gauge as described in Ref.~\cite{HefReiCam12}}
\label{fig3}
\end{figure}
By these relations the gap equation (\ref{12}) and the ghost DSE (\ref{14}) reduces to the ones in Coulomb gauge in the absence of the background field \cite{HefReiCam12}
\be
\label{21}
\omega^2 (\vp) = \vp^2 + \chi^2 (\vp) \, ,
\ee
\begin{gather}
\label{ghostform}
  d^{-1}(\vec{p}) = \frac{1}{g}-I_d(\vec{p})\,,\\
 I_d (\vp) = N_c \int \frac{\dd^3 q}{(2 \pi)^3} \bigl[ 1 -  (\hat{\vp} \cdot \hat{\vq})^2 \bigr] \frac{d(\vp - \vq)}{(\vp - \vq)^2} \frac{1 }{2 \omega (\vq)} \nonumber\,.
\end{gather}
Here $d(\vp)$ is the ghost form factor, defined by
\be
G(\vp) = \frac{d(\vp)}{g \vp^2}\,,
\ee
and we have replaced the full ghost-gluon vertex $\Gamma$ by the bare one $\Gamma_0$, which is known to be a good approximation, see Ref.~\cite{CamRei12}.
Lattice calculation \cite{BurQuaRei09} of the gluon propagator in Coulomb gauge show that the gluon energy can be nicely fitted by Gribov's formula \cite{Gribov78}
\be
\label{22}
\omega (\vp) = \sqrt{\vp^2 + M^4/\vp^2} \, .
\ee
A full self-consistent solution of the gap equation (\ref{21}) and the ghost DSE (\ref{ghostform}) reveals that $\omega (\vp)$ contains in addition sub-leading UV-logs,
which on the lattice are found to be small. Using Gribov's formula (\ref{22}) for $\omega (\vp)$ and solving the gap equation (\ref{21})
for $\chi (\vp)$ yields 
\be
\label{23}
\chi (\vp) = M^2 / |\vp| \, ,
\ee
which is indeed the correct IR-behavior obtained in a full solution \cite{HefReiCam12} of the coupled ghost DSE and gap equation show in Fig.~\ref{fig3} but which misses the sub-leading
UV-logs.

\section{The effective potential}

As explained in Sec.~\ref{sectionII} the constant background field residing in the Cartan algebra can serve as order parameter for
confinement when it is directed along a compactified dimension. Choosing $\va = a \ve_3$ and compactifying the 3-axis to a circle with 
circumference $L$, the shifted momentum (\ref{19}) becomes 
\be
\label{24}
\vp^\sigma = \vp_\perp + \lk p_n - \sigma a \rk \ve_3 \, , \quad p_n = 2 \pi n/L \, ,
\ee
where $\vp_\perp$ is the projection of $\vp$ into the 1-2-plane and $p_n$ is the Matsubara frequency. In the Hamiltonian approach the 
effective potential of the constant background field is given by the energy density in the state minimizing $\vev{ H }_a$
under the constraint $\vev{A }_a = a$ \cite{WeinbV2}. Using the gap equation one finds for the energy density per transversal degree of freedom $\langle H \rangle_a / (2 V)$
(V is the spatial volume) in the present approximation
\be
\label{25}
e (a, L) = \sum_\sigma \frac{1}{L} \sum^\infty_{n = - \infty} \int \frac{\dd[2] p_\perp}{(2 \pi)^2} \lk \omega \lk \vp^\sigma \rk - 
\chi \lk \vp^\sigma \rk \rk \, .
\ee
By shifting the summation index $n$ one verifies the periodicity
\be
\label{26}
e \lk a + 2 \pi/L , L\rk = e (a, L) \, ,
\ee
which is a necessary property for the effective potential of the confinement order parameter by center symmetry, cf.~Eq.~(\ref{3}). 
Neglecting $\chi (\vp)$ 
Eq.~(\ref{25}) gives the energy of a non-interacting Bose gas with single-particle energy $\omega (\vp)$. This quasi-particle
picture is a consequence of the Gaussian ansatz (\ref{10}) for the wave functional. The quasi-particle energy $\omega (\vp)$ is,
however, highly non-perturbative, see for example Eq.~(\ref{22}). The curvature $\chi (\vp)$ in Eq.~(\ref{25}) arises from the Faddeev-Popov determinant
in the kinetic part of the Yang--Mills Hamiltonian (\ref{5}).
\begin{figure}[t]
\includegraphics[width=0.8\linewidth]{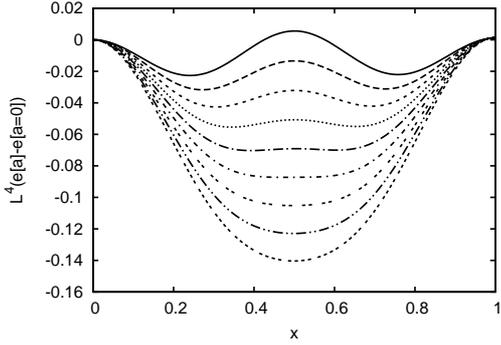}
 \caption{The energy density (\ref{25}) as a function of $x=a L/(2\pi)$. $L^{-1}$ was varied from $260$ to  $280$ MeV (from bottom to top).}
\label{fig2}
\end{figure}

In certain limiting cases and for $0 \leq a L / 2 \pi \leq 1$ the energy density (\ref{25}) can be calculated analytically. Neglecting $\chi (\vp)$ and assuming the perturbative
expression for the gluon energy $\omega (\vp) = |\vp|$ one finds from (\ref{25}) the Weiss potential originally obtained in \cite{Weiss:1980rj}
\be
\label{27}
e_\text{UV} (a, L) = \frac{4}{3} \frac{\pi^2}{L^4} \lk \frac{a L}{2 \pi} \rk^2 \lk \frac{a L}{2 \pi} - 1 \rk^2 \, .
\ee
Neglecting $\chi (\vp)$ and using the infrared expression for the gluon energy
$\omega (\vp) = M^2 / |\vp|$ (see Eq.~(\ref{22})), one obtains
\be
\label{28}
e_\text{IR} (a, L) = 2 \frac{M^2}{L^2} \left[ \lk \frac{a L}{2 \pi} \rk^2 - \frac{a L}{2 \pi} \right] \, .
\ee
This expression drastically differs from the Weiss potential (\ref{27}): While $e_\text{UV} (a, L)$ is minimal for $a = 0$, the minimum of
$e_\text{IR} (a, L)$ occurs at $a = \pi / L$ corresponding to a center symmetric ground state. Accordingly $e_\text{UV} (a, L)$ yields for the Polyakov
loop $\langle P \rangle = 1$ while $e_\text{IR} (a, L)$ yields $\langle P \rangle = 0$.

Obviously, the deconfinement phase transition is related to a change of the effective potential from its infrared behavior $e_\text{IR} (a, L)$
(\ref{28}) to its UV-behavior $e_\text{UV} (a, L)$ (\ref{27}). To illustrate this let us approximate the gluon energy $\omega(\vp)$
(\ref{22}) by 
\be
\label{29}
\omega (\vp) \approx |\vp| + M^2/|\vp| \, .
\ee
\begin{figure}[t]
 \includegraphics[width=0.8\linewidth]{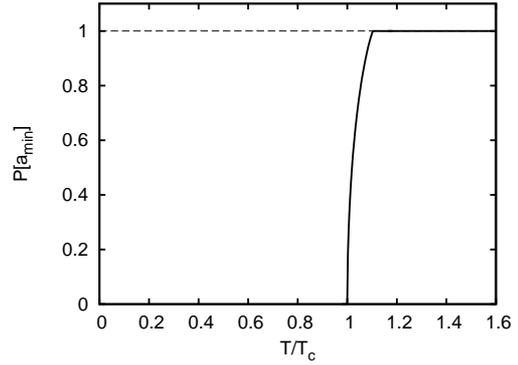}
\caption{The Polykav loop $\vev{P[a]}$  evaluated at the minimum $a = a_\text{min}$ of the full effective potential shown in Fig.~\ref{fig2} as a function of $T/T_c$.}
\label{fig4}
\end{figure}
This expression agrees with the Gribov formula (\ref{22}) in both, the IR and UV but deviates from it in the mid-momentum regime, which
influences the deconfinement phase transition. With $\omega (\vp)$ given by Eq.~(\ref{29}) and with $\chi (\vp) = 0$
the energy density (\ref{25}) becomes
\begin{gather}
\label{30}
e (a, L)  = e_\text{IR} (a, L) + e_\text{UV} (a, L) = \frac{4}{3} \frac{\pi^2}{L^4} f \lk \frac{a L}{2 \pi} \rk\,, \\
f (x)  =  x^2 (x -1)^2 + c x (x - 1) , \, c = \frac{3 M^2 L^2}{2 \pi^2} \nonumber \,.
\end{gather}
For small temperatures $L^{- 1}$, $e_\text{IR} (a, L)$ dominates and the system is in the confined phase. As $L^{- 1}$ increases the center symmetric
minimum at $x = 1/2$ eventually turns into a maximum and the system undergoes the deconfinement phase transition. In the deconfined phase 
$f (x)$ has two degenerate minima and, starting in the deconfined phase, the phase transition occurs when the three roots of $f' (x)$ degenerate.
This occurs for $c = 1/2$, i.e. for a critical temperature
\be
\label{31}
T_c = L^{- 1} = \sqrt{3} M / \pi \, .
\ee
With the lattice result $M = 880$ MeV this corresponds to a critical temperature of $T_c \simeq 485$ MeV, which is much too high. This, of course,
is not surprising given the approximation used to arrive at (\ref{31}), i.e. neglecting the ghost loop $\chi (\vp)$ and approximating
the Gribov formula by Eq.~(\ref{29}). Using the correct Gribov formula (\ref{22}) instead of the approximation (\ref{29}) only slightly reduces the critical temperature to $T_c \simeq 432$ MeV. It is the neglect of the curvature $\chi(\vp)$ which pushes the deconfinement phase transition to higher temperatures as can be easily seen: From the gap equation (\ref{21}) follows that in the deep IR $\omega(\vp)$ (\ref{22}) approaches $\chi(\vp)$ (\ref{23}). Therefore neglecting $\chi(\vp)$ in Eq.~(\ref{25}) increases the contribution of the confining part $e_\text{IR}(a)$ (\ref{28}) (relative to that of the deconfining part $e_\text{UV}$ (\ref{27})) and thus pushes the deconfinement phase transition to higher temperatures as we will also explicitly see below.

\section{Numerical Results}

We now turn to a full numerical evaluation of the effective potential (\ref{25}) using for $\omega (\vp)$
and $\chi (\vp)$ the numerical solution of the variational approach in Coulomb gauge obtained in Ref.~\cite{HefReiCam12} by solving the gap equation (\ref{21}) and the ghost DSE (\ref{ghostform}). The results for $\omega(\vp)$ and $\chi(\vp)$ are shown in Fig.~\ref{fig3}. With these results one finds from Eq.~(\ref{25}) the effective
potential shown in Fig.~\ref{fig2}. From this potential one extracts a critical temperature for the deconfinement phase transition of $T_c \simeq 269$ MeV,
which is close to the lattice predictions of $T_c = 290$ MeV. 
Let us also mention that if one uses for $\omega (\vp)$
the Gribov formula (\ref{22}) and in accord with the gap equation (\ref{21}) for $\chi (\vp)$ its infrared expression (\ref{23}) one finds a 
critical temperature of $T_c \simeq 267$ MeV, which is only slightly smaller than the value $T_c \simeq 269$ MeV obtained above with the full numerical solution for $\omega (\vp)$ and
$\chi (\vp)$. This shows that it is indeed the infrared part of the curvature $\chi(\vp)$ (neglected in Eq.~(\ref{30}), but fully included in Eq.~(\ref{25}) and Fig.~\ref{fig2}), which is crucial for the critical
temperature. In view of the ghost dominance in the IR this is not surprising. Fig.~\ref{fig4} shows the Polyakov loop $P[a]$ calculated from the minimum $a_\text{min}$ of the potential (\ref{25}) shown in Fig.~\ref{fig2}. At the phase-transition $P[a_\text{min}]$ rapidly changes from $P=0$ to $P=1$.

The value $T_c =269$ MeV obtained above from the full effective potential, Fig.~\ref{fig2}, is also close to the range of critical temperatures $T_c = 275 \ldots 290$ MeV obtained in Ref.~\cite{HefReiCam12} from the grand canonical ensemble of Yang--Mills theory in Coulomb gauge. It is however not surprising that the critical temperatures found in Ref.~\cite{HefReiCam12} differ somewhat from the value obtained in the present paper. The reason is that in the approach of Ref.~\cite{HefReiCam12} an additional approximation is made by using a singe particle ansatz for the density matrix. Such an approximation is not necessary in the present approach. In this respect the present approach is superior over the variational 
treatment of the grand canonical ensemble given in Ref.~\cite{HefReiCam12}.

In Ref.~\cite{Braun:2007bx} the Polyakov loop potential was calculated from a functional renormalization group flow equation approach using the Landau gauge ghost and gluon propagators as input.
For the gauge group SU$(2)$ a critical temperature of $266$ MeV was obtained, which compares well with our result of $269$ MeV.

In the present approach the deconfinement phase transition is entirely determined by the zero-temperature propagators, which are defined as 
vacuum expectation values. Consequently, the finite-temperature behavior of the theory and, in particular, the dynamics of the deconfinement
phase transition must be fully encoded in the vacuum wave functional, as should be clear from the considerations of Sec.~\ref{sec3}. The results obtained above are encouraging for an 
extension of the present approach to full QCD at finite temperature and baryon density. 

\section*{Acknowledgement}

One of the authors (H.R.) acknowledges useful discussion with J.M.~Pawlowski, M.~Quandt and G.~Burgio. We also thank D.~Campagnari and P.~Watson for a critical reading of the manuscript and useful comments. This work was supported by DFG under contract DFG-RE856/9-1 and
by BMBF under contract 06TU7199.

\end{document}